\documentclass[sigconf,nonacm]{acmart}
\renewcommand\footnotetextcopyrightpermission[1]{} % removes footnote with conference information in first column
\AtBeginDocument{%
  }

\setcopyright{none}
\copyrightyear{}
\acmYear{}
\acmDOI{}
\usepackage{xkeyval}
\usepackage{hyperref}

\newcommand{\revision}[1]{\textcolor{black}{#1}}%

\begin{document}


\title{Wasted large language models: A life cycle thinking approach}
%\title{Wasted LLMs: A life cycle thinking approach}
%\title{Stop wasting LLMs: A life cycle thinking approach}

%\title{Life cycle thinking and waste hierarchy for large language models}
%\title{Waste hierarchy for ML models: Reducing the climate impact of LLMs through life cycle thinking}
%\title{Lifecycle thinking for LLMs: Reducing waste of models and software}

\author{Erik Johannes Husom}
\affiliation{%
  \institution{SINTEF Digital}
  \city{Oslo}
  \country{Norway}}
\email{erik.johannes.husom@sintef.no}

\author{Maria Emine Nylund}
\affiliation{%
  \institution{SINTEF Digital}
  \city{Oslo}
  \country{Norway}}
\email{maria.nylund@sintef.no}

\author{Ophelia Prillard}
\affiliation{%
  \institution{SINTEF Digital}
  \city{Oslo}
  \country{Norway}}
\email{ophelia.prillard@sintef.no}


% Effort has been made to increase the energy efficiency of these models, however these do not reduced the overall consumption due to rebound effects such as Jevons paradox stating that increased effienciency lead to increased use. There is therefore a need for ...
\begin{abstract}
Large Language Models (LLMs) are machine learning (ML) models that have an increasingly large carbon footprint through their development and use. Efforts to increase the energy efficiency of these models have not translated into reduced consumption due to rebound effects such as Jevons Paradox - that increased efficiency drives increased use. There is therefore a need for additional measures to solve this problem.

We suggest that one possible way forward is to use life cycle thinking, and view LLMs as products that can become waste. With this perspective, we investigate the potential of the waste hierarchy from the EU's Waste Framework Directive, which suggests five different measures for how to manage waste: prevention, reuse, recycling, recovery, and disposal. 
We examine how these measures can inform and motivate new types of thinking and approaches to reducing LLM waste and their environmental impact in general.

Applying the waste hierarchy to LLMs highlights that preventing waste is essential for reducing the models' environmental impact, mainly because it reduces the need for training new models. Prevention can be achieved through many existing methods for reusing, "recycling", and "recovering" LLMs. Additionally, disposal can be important both for saving energy and for keeping a considerate attitude to the resources being spent on training LLMs. We also call to attention that prevention of unnecessary use of LLMs carry huge potential for lowering the climate impact of the models.

%While originally intended to serve as a framework for material products, we explore how it can be applied to immaterial products such as software, and LLMs in particular. 
%We describe how we can view ML models as software and products, and how to define when such products become waste. 

%The main elements of the waste hierarchy that are applicable to LLMs are prevention, reuse, and recycle. The concepts of recovery and disposal are less suited for immaterial products like software, but can still serve as inspiration for reassessing our attitude to the lifespan of LLMs and software in general.

\end{abstract}



\maketitle

\section{Introduction}

Large Language Models (LLMs), trained to generate text such as natural language and computer code, is one of the most popular forms of machine learning (ML) models. The proposed benefits of LLMs are many, and typically involve increased productivity and efficiency in a range of work tasks \cite{al2024enhancing}, especially in software development \cite{mohamed2025impact}. However, LLMs and other forms of large ML models have a considerable carbon footprint, requiring enormous amounts of energy to train and use at scale, and the climate impact is growing \cite{devriesGrowingEnergyFootprint2023}. 

%The general-purpose nature of LLMs makes them applicable in many sectors, and 
%and also means that researchers of many fields can work together to make improvements, instead of working separately on task-specific models. 

Because of this, we need actions to move towards a more environmentally sustainable approach to AI. Research fields like Green AI \cite{schwartzGreenAI2019} bring focus to methods for increasing energy-efficiency of the development and deployment of AI, especially regarding improved training strategies for machine learning models \cite{verdecchiaSystematicReviewGreen2023}. Energy efficiency is, however, not enough, as Jevons paradox typically leads to increased total consumption \cite{luccioni2025efficiency}. Additionally, the training methods for LLMs are advancing rapidly, introducing improved ways of pre-training foundational models\cite{velu2024llm, tirumala2023d4, bergsma2026power}, and adding additional and more complex steps of both post-training \cite{djuhera2026fixing, kumar2025llm} and so-called mid-training \cite{tu2025survey, mo2025mid}. 
%Data curation is also a crucial part of obtaining high performance for LLMs. 
Such rapid improvements \revision{create} incentives for increasingly frequent training and retraining of LLMs.

The increasing number of people working to improve LLMs, combined with a higher-than-ever demand for the models, is another factor that results in a large number of LLMs being trained. \revision{There} are many frontier labs, such as OpenAI, Anthropic, Google, and xAI, that offer proprietary LLMs as services, some of which are reaching record breaking numbers of users \cite{huChatGPTSetsRecord2023}. Additionally, there exists a large ecosystem of open-weight LLMs \cite{kukreja2024literature}, which are downloadable models that can be deployed and used by anyone. With so many people developing LLMs, it leads to questions regarding how we treat LLMs as products, how quickly they are deprecated, and whether the costs of training numerous LLMs is worth the benefits, if the lifespans of models are shortened by the constant stream of new ones.

%how many models are actually produced each year (or month), and whether we can reduce the climate impact by thinking differently around how models are used and deprecated.

%The first open-weights models with capabilities of producing human-like text were the infamous leak of Facebook's Llama model. Since then, many other labs offer their models openly, contributing to making the technology accessible. While one can argue that this democratizes the access to LLMs, there is another side of the coin: Unnecessary deprecation of models.

One way of gaining a deeper understanding of how we can get the most benefits out of products at minimal cost and waste is to use Life-cycle thinking (LCT) \cite{mazzi2020introduction}. A part of LCT is to consider how we deal with waste from products and services, and this type of thinking has led EU to introduce a waste management hierarchy \cite{directive2008directive}, consisting of the five steps of prevention, reduction, reuse, recycle, and responsible disposal. While software in general seems to be outside the scope for the proposed hierarchy, we may discover novel approaches to reducing AI's climate impact by reassessing how we use and discard software.

Our contribution is to describe and discuss how life cycle thinking can be applied to the development and deployment of LLMs, by using EU's waste hierarchy as a framework. We use the concepts of this framework to highlight how it can help us reduce unnecessary training costs and improve reuse of existing artifacts, and what the weaknesses of LLMs are in the context of life cycle thinking.

The paper is structured as follows: In Section \ref{sec:background}, we present background on life cycle thinking, the waste hierarchy, and how machine learning models can be viewed as software and products. In Section \ref{sec:waste-hierarchy-for-llms} we describe how LCT and the waste hierarchy can serve as a framework for improving environmentally responsible development and deployment of LLMs, followed by a discussion in Section \ref{sec:discussion} and conclusion in Section \ref{sec:conclusion}.
%Section \ref{sec:case-study} contains a case study where we use data collected from the HuggingFace platform to assess the current landscape of LLM releases, before we conclude in Section \ref{sec:conclusion}.

\vspace{-0.3cm}

\section{Background}
\label{sec:background}

To investigate how we can apply waste management principles on LLMs, we will first introduce life cycle thinking and EU's waste hierarchy, and then present perspectives on how we can view machine learning (ML) models as software.

\subsection{Life cycle thinking and the waste hierarchy}

Life cycle thinking (LCT) \cite{mazzi2020introduction} means to consider the full life cycle of products and services when assessing their environmental impact. This means taking into account not only the costs related to production and use, but also what happens to a product after it is done with its useful lifetime. A product enters this final part of its life cycle when it is no longer of use or valuable for its original intended purpose, and we may define it as "waste".

\begin{figure}
    \centering
    \includegraphics[width=0.7\linewidth]{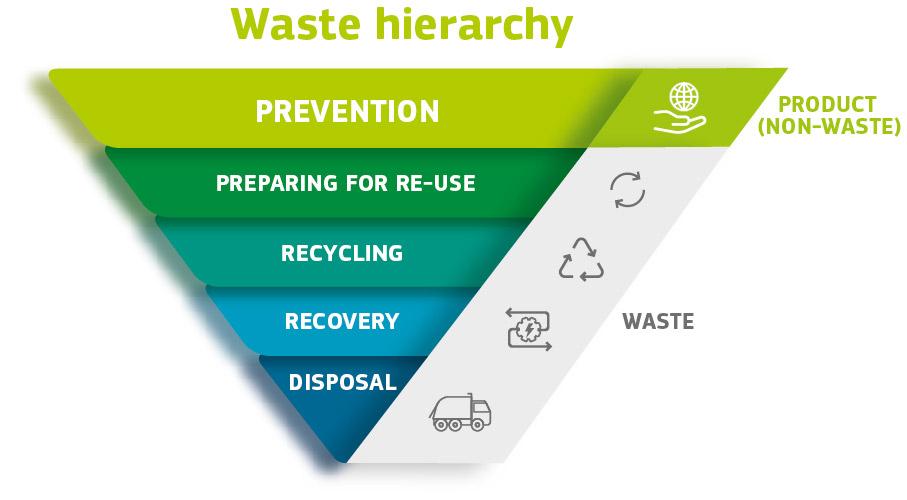}
    \caption{The waste hierarchy, as proposed by EU's Waste Framework Directive. Source: \href{https://environment.ec.europa.eu/topics/waste-and-recycling/waste-framework-directive_en}{https://environment.ec.europa.eu/topics/waste-and-recycling/waste-framework-directive\_en}}
    %\caption[The waste hierarchy, as proposed by EU's Waste Framework Directive.]{The waste hierarchy, as proposed by EU's Waste Framework Directive.\protect\footnotemark}
    \label{fig:waste-hierarchy}
\end{figure}

LCT has served as guidance on EU's waste policy and regulation \cite{europeancommission2010lifecyclewaste}. The European Comission introduced in 2008 the Waste Framework Directive (WFD) \cite{directive2008directive}, which is a legal framework for preventing and managing waste in the EU. The framework established a "waste hierarchy", as shown in Figure \ref{fig:waste-hierarchy}, giving a prioritized list of five measures taken to improve environmental outcome related to waste (the definitions of each measure is taken from the directive itself):

\begin{enumerate}
    \item \textbf{Prevention.} Measures taken before a given product or material has become waste.
    \item \textbf{Preparing for re-use.} Enabling products or parts of products to be used again for its original purpose.
    \item \textbf{Recycling.} Reprocessing waste into new products or materials.
    \item \textbf{Recovery.} Processes that enables recovery of components, material, or energy from the waste.
    \item \textbf{Disposal.} Processes for destruction, storage, or discarding of waste that does not involve recovery of materials, such as \revision{depositing} on to land, releasing into the environment, incineration, or permanent storage.
\end{enumerate}

%\footnotetext{Source: \href{https://environment.ec.europa.eu/topics/waste-and-recycling/waste-framework-directive_en}{https://environment.ec.europa.eu/topics/waste-and-recycling/waste-framework-directive\_en}}

The waste hierarchy is in general targeted at material products, but in this paper, we investigate its potential for being applied to immaterial products such as software, and in particular machine learning models such as LLMs.

\subsection{Machine learning models as software}

Software can be defined as instructions given for a computer to follow. ML models such as LLMs are software in the way that the models contain low-level instructions that are executed by a computer. The models are in essence a set of numbers that are multiplied or added together in a certain manner given an input, producing a certain output. 

However, ML models are unlike computer code, which is the traditional form of software. Code comes into existing by having someone write explicit instructions in any given programming language. ML models need several elements to be produced: 1) Training data from which a certain task should be learnt to perform, and 2) an ML algorithm that specifies how certain parameters should be optimized in order for the resulting model to perform the task. The output of this is not human-readable computer code, but low-level instructions on how to process certain input by multiplying and adding a set of numbers in a very specific way. The computer scientist Andrej Karpathy popularized this understanding in 2017, when he argued that software has evolved beyond computer code (which he calls "Software 1.0") \cite{karpathy2.0}. Now, artificial neural networks (ANNs), a form of machine learning, represent "Software 2.0", where the instructions for the computer is encoded as weights of a neural network. 

The differences between these two types of software have implications for the potential of the waste hierarchy as a method for reducing the environmental impact of LLMs. Both computer code and LLMs are products, in the sense that they are artifacts created and used by humans. Seeing software as a product is the basis for how we can define "software waste".

%\subsection{Machine learning models as products}

%The implications of that product being immaterial will become clear during the discussion.

%Distingiushing between these two forms of providing instructions to a computer is important, because it highlights certain

%In 2025, Karpathy expanded the theory and introduced Software 3.0, where the prompts given to LLMs

%Already, we can discern important differences in how these two types of software work, and how they can be adjusted.

%The ongoing climate crisis has led to a lot of life-cycle assessments and estimation of carbon emissions for a lot of products and services.

\section{Waste hierarchy for LLMs}
\label{sec:waste-hierarchy-for-llms}

%Generative AI models, as they are currently developed and deployed, parts from the existing paradigm within software eingeering. Software is expensive and time-consuming to produce
To apply the waste hierarchy to LLMs, we first need to define what we mean by LLMs becoming waste, and explain why reducing LLM waste can reduce the environmental impact of LLMs. Then we can proceed to explore how measures from EU's Waste Framework Directive can be useful.

\subsection{When do LLMs and software become waste?}

Software is not material, and the concept of "waste" is therefore not clearly defined or understood in this setting. Sedano et al. described a taxonomy of waste in software development, and used the definition of waste as "any \textit{activity} that consumes resources but creates no value for the customer [emphasis added]" \cite{sedano2017software}, a definition shared with other papers on the same topic \cite{ikonen2010exploring, alahyari2019exploratory}. In this paper, however, we are not looking at waste from processes, but a product that becomes waste.
%given the context of EU's waste hierarchy, we define "waste" not as an activity, but as a product. 
EU's Waste Framework Directive defines "waste" as "any substance or object which the holder discards or intends or is required to discard" \cite{directive2008directive}. We can adjust this definition to be applicable to LLMs, and software in general, by simply replacing "substance or object" with "product". An LLM becomes waste when it is discarded and no longer used.

%Because of LLMs, and software in general, can be copied
The immaterial nature of LLMs and software makes them trivial to copy and reuse. While we can never definitely say that a given model has fully become waste, since a copy of it may be running somewhere, we can observe that when new models are released with better performance and/or lower cost, older models are deprecated because there are few or no reasons to use them when something better is available. 
%LLMs are quickly deprecated in the sense that new models are frequently trained, resulting in better performance on benchmarks for the same inference cost.

\subsection{Reducing climate impact by reducing LLM waste}

LLMs have an environmental impact through the energy usage of:

\begin{enumerate}
    \item developing (data preprocessing, experimentation, training, and validation), 
    \item storing, 
    \item moving (transmitting over network, for example through uploading and downloading), and
    \item using (running inference) the models. 
\end{enumerate}

\revision{We can reduce the environmental impact of LLMs if we can reduce the need for any of these four processes, and thereby their energy consumption. Given that discarded software does not occupy space, carry no risk of releasing harmful substances, and in general can be disposed of without leaving any residue, the concept of waste management for improved environmental sustainability is motivated mainly by \textit{how it can help us avoid unnecessary production of new models and better reuse of existing models}. This will in turn reduce the energy consumption of 1) developing, 2) storing, and 3) moving new models. The effect on the impact of 4) using models depends on whether a larger number of available models increases, decreases, or have no effect on the usage of LLMs, and is left out of the discussion of this paper.}

\revision{Reducing the actual use of models is an important method for reducing environmental impact of LLMs in general, but it is a separate discussion, as we in this paper discuss how we can improve our use of already existing models, assuming that they are already produced and will be used. We expand briefly on this subject in Section \ref{sec:prevention-of-use}.}

\revision{Other climate impacts, such as those from the hardware production required for developing and running LLMs \cite{falkFLOPsFootprintsResource2025a}, land use of data centers \cite{obringer2021overlooked}, and water use for cooling of data centers \cite{siddik2021environmental} are also important in general. However, waste management of LLMs, as proposed in this paper, has its first-order effects on the four processes mentioned above, making hardware-related impacts secondary. Additionally, the Waste Framework Directive already applies to material products, and we therefore consider the secondary impacts outside the scope of this paper.}

%those secondary effects outside the scope of this paper, as the waste hierarchy already applies to material products. However, reduced operational energy usage caused by LLMs will in turn lead to reduced need for hardware, cooling, and other related impacts.}

%In this paper, however, we focus on what we may call the \textit{first-order} environmental impacts of LLMs: The impacts that are caused directly from the processes of developing and using the product, excluding the impact from making or maintaining the equipment used in those processes. }

%Since it can be unfamiliar to think about LLMs as a product in itself, it can be helpful to consider an analogy to understand how we treat them in this paper: Let us say that the product under consideration is a car. The first-order environmental impact is the energy and emissions from manufacturing the car and from driving it. We do not include the environmental impact from the 

\subsection{Waste management measures applied to LLMs}

%The last part of the hierarchy, "Disposal", is therefore considered irrelevant, and we focus only on the four first parts.

In this section, we will go through each of the five measures from EU's waste hierarchy, and describe how the measures can help us reduce waste of LLMs.

%For LLMs, there are obvious benefits from each of these stages:

\subsubsection{Prevention and reuse}

%Prevention and reuse are tightly connected. 
The relevant parts of the Waste Framework Directive's definition of "prevention" include measures that reduce "the quantity of waste, including through the re-use of  products or the extension of the life span of products" \cite{directive2008directive}. 
Reusing LLMs is arguably the most obvious way of avoiding that the models become waste, and extending the life span of products enables more reuse. We will therefore discuss prevention and reuse together.

Reuse of software is happening on an extremely large scale through the use of libraries, frameworks, compilers and operating systems. It is arguably the whole foundation of software and computing systems, since subroutines are reusable components that are combined for various purposes. LLMs can be copied an unlimited number of times, and reuse is therefore a trivial way of amortizing the energy spent on developing, storing, and moving the models.

Mistral, which is one of the few AI labs that has published a full life-cycle assessment of one of their models, specifically note that one of the key indicators that are needed to control the environmental impact of AI is "the ratio of total inference to total life-cycle impacts", because we should "ensure that models’ training phases are amortized, and not wasted" \cite{OurContributionGlobal}.

Since LLMs cannot be broken or worn down like material products, the greatest barrier for reuse is arguably deprecation, because new and improved models are released. One approach is to shift the focus from "performance at all costs" to sufficiency and what can be considered good enough. Slower turnover of models in LLM-driven applications will also lead to more stable software, because it requires less frequent adaptation to new models and capabilities.

One may also consider approaches that "augment" LLMs through external information sources as ways of reusing the models while still achieving improved performance. Retrieval augmented generation (RAG) \cite{lewis2020retrieval} is an example of this, where LLMs can access knowledge bases to improve their ability to provide text that is factually correct. Another interesting approach is the concept of "agent skills", introduced by Anthropic \cite{IntroducingAgentSkills}, which is a way of providing reusable instructions to an LLM. In essence, it allows for reuse of prompts and context, and through the refinement of agent skills, one may reduce the amount of inference needed to complete a task with an LLM.

Another approach for increasing the life span of LLMs is to explore ways of extending, modifying, and otherwise adapting existing LLMs to stay useful for longer. We will discuss such measures in the next section.

%Preventation: Do not use LLMs if not necessary. It can however be difficult to assess whether what is more wasteful: Asking a language model, or doing multiple searches with traditional tools. Factoring in the learning effects, etc, the whole questions becomes very difficult, also given that the energy use is not very transparent.

%Reuse: This is the most substantial part, with many potential ways of reusing.
%reusing for same purpose

%\subsubsection{Reuse}
%\label{sec:reuse}

\subsubsection{Recycle and recovery}

Recycling means "any recovery operation by which waste  materials are reprocessed into products", while recovery is more generally that waste continues to serve "a useful purpose" \cite{directive2008directive}. We cover these two measures together, as recycling can be considered a subset of recovery.

There are many existing methods for what can be considered as recycling of LLMs, which we define as any method that extend or adapt existing LLMs. One example is fine-tuning, which means to use an already trained model as a foundation for further training on another dataset than it was originally trained on \cite{parthasarathy2024ultimate}, usually for making the model more suitable for a specific task. 
Low-rank adaptation (LoRa) \cite{hu2022lora} is a form of model fine-tuning, where the original model stays fixed, but the fine-tuning is done by only training a small set of parameters that are added on to the original model. LoRa then functions as an adapter. This has inspired the creation of reusable libraries of LoRa adapters \cite{ostapenko2024towards}, which can make it even easier to reuse and recycle LLMs. \revision{In many cases, fine-tuning will in practice serve as a form of reuse rather than recycling, as it is often applied to models that are not (yet) waste. It depends on whether the original model is one that would otherwise be discarded.} \revision{For recycling,} LLM merging \cite{tam2024llm, fu2025training} is another relevant approach, which can be used to obtain improved capabilities of LLMs by combining existing models instead of training new ones.

Additionally, there are techniques such as quantization and pruning \cite{kuzmin2023pruning}, allows for adapting and optimizing LLMs to run on resource-constrained devices, meaning one does not necessarily need to train specific models matching the resources available. 

Software as computer code is trivial to recycle, because parts of code, whether that is snippets or modules, typically can easily be included in new pieces of software. This highlights a weakness of LLMs as a new paradigm for software: They require much more complex techniques to recycle than traditional software.

%Recycle: This process has no clear equivalent in the ML model world, as software does not need to be composted or recovered.

%\subsection{Recovery}
%Recovery is understood as recovering certain elements of the product that can be . No clear equivalent.
%What would recovery look like for LLMs?
%Compare with software.
%Maybe the concept of recovery for material waste can inspire new ways of exploring opportunities 

\subsubsection{Disposal}

LLMs, and software in general, have no need of disposal in the traditional sense, as the artefacts can simply be deleted without having any residue. However, file storage consumes a significant amount of energy, and deletion of unused data is therefore beneficial for reducing climate impact \cite{al2022exploring, mersico2024challenges}. The storage sizes of an LLM can typically be from a few to several hundreds of GBs \cite{OptimizingLLMsSpeed}, meaning there can be substantial savings by actually disposing of unused LLMs.

At the same time, the ease of digital deletion highlights the throwaway mentality this breeds. LLMs can be disposed of at no cost, it disregards the energy and time it took to produce the models, and can lead us to be inconsiderate, especially given that the resource consumption of developing LLMs is largely hidden from both developers and users.

%While it may seem like this makes it irrelevant to talk about disposal in this context, 

%However, the ease of so

%\section{Case study: Text generation models at HuggingFace}
%\label{sec:case-study}
%
%Studying use of LLMs in the wild is difficult, because it would rely on a high degree of surveillance to observe how much and for what the models are actually used. Downloadable models give the great benefit of distributed use, but that also entails that it is challenging to get a clear picture of the lifespan of models.
%
%HuggingFace is the largest platform for sharing machine learning models, and provides statistics on how many times any given model has been downloaded.
%
%Number of downloads does not say anything about how many times it is used.
%
%\begin{figure}
%    \centering
%    \includegraphics[width=0.9\linewidth]{figures/monthly_releases.pdf}
%    \caption{Caption}
%    \label{fig:monthly_releases}
%\end{figure}
%
%Combined with recent data from OpenRouter, we can get a better picture of what models are actually used the most by certain groups of users.
%
%However, we don't know anything about the overlap between those that download models and those that choose to use models through an API.
%
%The platform cannot provide data on inference usage, but the numbers may imply how many models are being trained, and
%
%
%\remark{TODO: Finish subsection}

\subsection{Prevention of unnecessary use of LLMs}
\label{sec:prevention-of-use}

While we mainly discuss prevention in the context of preventing waste of LLMs as a product in this paper, it is worth noting that preventing unnecessary use of LLMs is an important contribution to reducing the environmental impact of LLMs. To make the distinction clear: On the one hand, LLMs can be wasted if we spend a lot of energy to train them, but then do not use them much before we feel the need to train a new one (this is the primary understanding of "LLMs as waste" in this paper). On the other hand, we waste a lot of energy if we use LLMs for tasks where we could have used a much more efficient tool. Given that LLM inference has a very large energy footprint due to large-scale deployment \cite{jegham2025hungry}, there is a huge potential for energy savings by considering more efficient tools for a wide range of tasks.

We refrain from going into a detailed description of use cases where LLMs are significantly more expensive to use than more traditional tools \revision{(for example, using a frontier LLM for generating boilerplate code when existing templates suffice)}. We will, however, note that there is a gap in human-computer interaction (HCI) on how to promote LLM usage and life cycle thinking that make users aware of the \revision{environmental} costs and what they can do about it. For example, when it comes to preventing unnecessary use of LLMs, there is nothing in the typical interface of LLM-powered chatbots that helps the user to avoid needless use of the LLM, other than the ability to choose a more efficient model or use a mode with "lower effort". On the contrary, there are lots of dark patterns being employed to steer the user towards using LLMs \cite{beignon2025imposing}.

\section{Discussion}
\label{sec:discussion}

Looking at the potential ways of reducing waste of LLMs, avoiding the training of new models is one of the main ways we can use the waste hierarchy to reduce climate impacts. Our overview of approaches for reducing the waste of LLMs is by no means exhaustive, but outlines a way of thinking that may allow for better "waste management" of LLMs.

Nevertheless, training new models is often required for \revision{improvement}, e.g., to understand the effect of new data curation methods and training approaches. New models may also be necessary to mitigate important limitations in existing models, such as bias \cite{ranjan2024comprehensive} and "hallucinations" \cite{perkovic2024hallucinations}, although one can also use "augmentation" strategies such as knowledge graphs or other external sources to reduce such weaknesses \cite{agrawal2024can}. Understanding when it is worth to train a new model or not is crucial to avoid waste of LLMs.

%Several incentives may be behind the search for better performance: Scientific curiosity, capital gain and increased market share, and in general attempting to solve problems.

To better understand how LLMs are deprecated and becomes waste, one needs to collect data that can give a better view of the lifespan of models, how much they are used, when their usage decline, and factors such as the relative improvements in performance and efficiency for each new model. Ana Trisovic \cite{trisovicShrinkingLifespanLLMs2026} studied the lifespan of LLMs in research, by tracking 62 LLMs during the period 2018-2025 by analysing 108 000 citing papers. The results shows that the lifespans of LLMs are getting shorter and shorter when measured on adoption and interest from academia, and it is mainly due to the competitiveness of the market. As Trisovic notes: "A model’s lifecycle is [...] less a function of its intrinsic properties than of the competitive landscape it enters." Future work should investigate these trends beyond the research community, and look at how LLMs are developed and deployed by both industry, practitioners, and consumers.

%\section{Limitations}
%\label{sec:limitations}

The value of LLMs is highly debated. While the models certainly can generate text at incredible speed, rebound effects of such as cognitive outsourcing, atrophying skills, and technical debt in computer code, makes it unclear how much net benefit the technology brings. This paper looks at LLMs simply as a product and discusses how to reduce waste of the models, without taking into account the actual value of the product. The current lack of understanding of long-term effects is an underlying limitation of all discussions around the costs of LLMs, as costs should always be weighed against benefits. If one considers LLMs in general as devoid of any net benefits, the whole discussion of waste of LLMs is meaningless, as LLMs in themselves would then be considered waste.

\revision{In a time when model capabilities are increasing faster than ever \cite{epoch2025aicapabilitiesprogresshasspedup}, it seems unlikely that sufficiency-based approaches like reuse and recycling can gain traction. It is difficult, and perhaps unrealistic, to create effective incentives for this in a growth-based economy. However, there are many underlying incentives for being more climate-conscious: avoid the worst effects of climate change, save resources for more critical uses, and reduce destruction of nature.}

\revision{The aim of this paper is not to develop guidelines that are realistically passed into law or mainstream practice in the short-term. Our proposed framework can, as a starting point, help developers and users of LLMs become more conscious about the resources spent on developing models. The framework highlights techniques for using what we already have, rather than spending a large amount of resources on developing new models. In short, it is a method for combatting throw-away mentality when it comes to LLMs. }

%serving the goal of improving LLMs performance on given tasks or capabilities. 

\vspace{-0.1cm}

\section{Conclusion and future work}
\label{sec:conclusion}

In this paper we have looked at how life cycle thinking and the waste hierarchy from EU's Waste Framework Directive can be applied to LLMs, and help us reduce the environmental impact of this technology. Thinking about LLMs as products that can become waste reframes the discussion around their development and deployment, and highlights how a more considerate approach can lead to less climate impact. We have analyzed the waste hierarchy's five measures of prevention, reuse, \revision{recycling}, recovery, and disposal, and suggested methods and perspectives on how to apply these concepts in the context of LLMs. This study motivates a deeper investigation into the lifespan of LLMs and the effect of the various measures we have proposed.

\vspace{-0.2cm}

%LLM-powered AI agents 
%agentic ai: reuse plans \cite{li2025plan}

\begin{acks}
The work has been conducted as part of the ENFIELD project (101120657) funded by the European Commission within the HEU Programme.

\revision{Special thanks go to Stine Vintervoll, senior sustainability advisor at Miljøfyrtårn, who inspired the idea for this paper through thoughtful discussions.}
\end{acks}

\bibliographystyle{ACM-Reference-Format}
\bibliography{main}

\end{document}